\documentstyle[preprint,aps,epsf]{revtex}
\tighten

\begin{document}

\draft

\title{Three-Nucleon Photodisintegration of $\bbox{^3}$He
}

\author{
R.~Skibi\'nski$^1$,
J.~Golak$^{1,2}$,
H.~Wita\l{}a$^1$,
W.~Gl\"ockle$^2$,
H.~Kamada$^3$
A.~Nogga$^4$.
}
\address{$^1$M. Smoluchowski Institute of Physics, Jagiellonian University,
                    PL-30059 Krak\'ow, Poland}
\address{$^2$Institut f\"ur Theoretische Physik II,
         Ruhr-Universit\"at Bochum, D-44780 Bochum, Germany}
\address{$^3$ Department of Physics, Faculty of Engineering,
   Kyushu Institute of Technology,
   1-1 Sensuicho, Tobata, Kitakyushu 804-8550, Japan}
\address{$^4$ Department of Physics, University of Arizona, Tucson,
              Arizona 85721, USA}

\date{\today}
\maketitle

\begin{abstract}
The three-nucleon photodisintegration of $^3$He has been calculated in the whole phase space using
consistent Faddeev equations  for the three-nucleon bound and scattering states. Modern
nucleon-nucleon
 and 3N forces
have been applied as well as different approaches to nuclear currents. Phase space regions
are localized where 3N force effects are especially large. In addition, semi-exclusive cross
sections for $^3$He$(\gamma$,N),
which carry interesting peak structures,
have been predicted.
Finally, some data for the exclusive 3N breakup process of $^3$He and its total
breakup cross section have been compared to theory.
\end{abstract}

\narrowtext

\section{Introduction}
\label{secIN}

The reactions $^3{\rm He}(\gamma, np)p$ and $^3{\rm He}(\gamma, pp)n$
have been studied
experimentally in the past below and above the pion threshold. In
this paper we only look to the energy regime below the pion
threshold. In~\cite{ref1}
and \cite{ref2}
these processes have been investigated
in relation to
a search for three-body absorption
mechanisms or to observe quasi-deuteron breakup.
We refer the reader for earlier studies to these
two references.
Quite a few pioneering theoretical studies have been performed by
Laget~\cite{ref3}
and applied to these reactions. These calculations are done
using a certain class of diagrams consisting of absorption mechanisms of
the photon at one, two and three nucleons and allowing for low orders rescattering
among the nucleons. Pioneering calculations in the
framework of Faddeev equations and based on S-wave spin dependent
separable potentials have been carried through
in~\cite{ref4}.
There cross
sections for the semi-exclusive processes
$^3{\rm He}(\gamma,$N) have been
determined.

 In a previous paper~\cite{Romek.2b} we investigated the two-body breakup
process of $^3$He ($^3$H) with
the aim to search for three nucleon (3N) force effects. We found that most of the existing data
supported qualitatively the predicted three nucleon force (3NF) effects
but new precise data would be very welcome
to challenge theory more
strongly.

Here we are mostly interested in kinematically complete 3N breakup
processes and shall employ
rigorous solutions of the Faddeev equations
consistently for the 3N bound state and the 3N continuum. Modern nucleon-nucleon (NN) and
3NF's will be used and mesonic exchange currents (MEC) will be employed
either explicitly or in the form of the Siegert approximation.
Both forms were previously used and described in~\cite{Golak.pdc.2000} to investigate the p-d
capture process.
The emphasis
of the present investigation is put on predicting those  regions of the 3N
phase space, where 3NF effects are especially pronounced. Of course,
this is  based on
present-day 3NF model. We should also remark   that  we have
 not yet included explicit  3N
electromagnetic current operators, which  are required by the
continuity equation and which may play a role. In case of the
Siegert approximation, however, some of them are automatically included.

In addition to the fully exclusive breakup cross section, we also present
theoretical predictions for the semi-exclusive processes
$^3{\rm He}(\gamma, p)$ and $^3{\rm He}(\gamma,n )$.
They show interesting peak structures based on a complex interplay of all
dynamical ingredients.

In relation to the two experimental investigations \cite{ref1} and \cite{ref2}
we shall show some related point geometry results but
unfortunately are unable to fully analyze those data.
This is due to insufficient access to the experimental details.
In the present study
 we shall also compare the theory to existing total 3N breakup
data on $^3$He and $^3$H
measured in the low energy region.
After finishing this work we heard of $^3$He breakup data~\cite{Debev},
which possibly might be analyzed in the future.

A very recent paper~\cite{Trento} also deals
with total
photodisintegration cross sections. There, besides studying 3NF effects,
 the emphasis was
placed on performing a benchmark between two totally different approaches:
 the Faddeev one in
momentum space  and a hyperspherical harmonic expansion method
in configuration space
combined with a Lorentz transform method. The results agreed quite
 nicely documenting the
present-day accuracy in treating these quite complicated processes
 numerically for 
certain types of nuclear forces and electromagnetic current operators.

We briefly describe our theoretical framework in Sec.~II and display
our results in Sec.~III. The summary is given in Sec.~IV.

\section{Theoretical Framework}
\label{secII}

We refer to~\cite{gloecklebook,Golak.PRC51} and~\cite{ref.Report}
for our general notation
and specifically to \cite{Romek.2b}
for the formalism of $^3$He photodisintegration. As is shown there, the
nuclear matrix element

\begin{eqnarray}
N_\tau^{\rm 3N} \equiv \langle \Psi_{\vec p \, \vec q}^{(-)}
\mid j_\tau (\vec Q ) \mid \Psi_{^3{\rm He}} \rangle .
\label{eq:N3N}
\end{eqnarray}
for 3N breakup of $^3$He can be written as

\begin{eqnarray}
N_\tau^{\rm 3N} =
\frac12  \langle \Phi_0 \mid ( t G_0 + 1 ) P \mid \tilde{U} \rangle ,
\label{eq:N3NN}
\end{eqnarray}
where $\mid \tilde{U} \rangle $ obeys the Faddeev-type integral equation

\begin{eqnarray}
\mid \tilde{U} \rangle  \ = \
( 1 + P ) j_\tau (\vec Q ) \mid \Psi_{^3{\rm He}} \rangle \ + \
\left( t G_0 P + \frac12 ( 1 + P ) V_4^{(1)} G_0 ( t G_0 +1) P \right)
\mid \tilde{U} \rangle .
\label{eq:Utilde}
\end{eqnarray}

We encounter in Eq.(\ref{eq:N3N}) the asymptotic relative momenta $\vec p$ and $\vec q$
  of the three final
nucleons attached to the 3N scattering state
$\langle \Psi_{\vec p \, \vec q}^{(-)} \mid  $
and the spherical component
$ j_\tau (\vec Q ) $
of the electromagnetic current operator.
Further, $ \langle \Phi_0 \mid $  is a properly
anti-symmetrized (in the two-body subsystem) free 3N state, $t$ the NN t-
operator, $G_0$ the free 3N propagator and $P$ the sum of a cyclical and
anticyclical permutation of 3 particles. Finally, $V_4^{(1)}$
is that part of a 3NF, which is symmetrical (like the NN $t$-operator) under exchange
of particles 2 and 3.
That Faddeev equation can
be solved rigorously in momentum space using a partial wave decomposition. Any
NN force (leading to $t$) and 3NF can be used.

The fivefold differential cross section for the complete $^3$He ($^3$H)
breakup is given as
\begin{equation}
{ {d^{\, 5} \sigma} \over { d \Omega_1 \, d \Omega_2 \, d S } } \ = \
{ 2 \pi^2 \alpha \over E_\gamma } \, { 1 \over 2 } \,
\sum_{M,m_1,m_2,m_3} \, \left( \mid N_{+1} \mid ^2 + \mid N_{-1} \mid ^2 \right)
\ \rho _{3N} ,
\label{eq:sigma5}
\end{equation}
where $\alpha$ is the fine-structure constant.
In order to avoid kinematical singularities, we represent the breakup cross
section along the kinematically allowed locus in the $E_1 $ - $E_2$
plane and use the arc-length $S$ along that locus (on which all
events have to lie for fixed $\Theta_1$, $\Phi_1$, $\Theta_2$ and $\Phi_2$)
to label the cross section.
Then the nonrelativistic phase-space factor is

\begin{equation}
\rho_{3N} = { { m_N^2 \ \vert {\vec p}_1 \vert \ \vert {\vec p}_2 \vert }
\over { \sqrt{ \vert 1 - { { {\vec p}_2 \cdot {\vec p}_3 }
\over { \vert {\vec p}_2 \vert ^2}                } \vert ^2
+ \vert 1 - { { {\vec p}_1 \cdot {\vec p}_3 } \over { \vert {\vec p}_1 \vert ^2 }
            } \vert ^2
             }
      }
            } ,
\end{equation}
where the momenta of the two detected nucleons
are denoted by ${\vec p}_1 $ and ${\vec p}_2 $, respectively, and the
nucleon mass by $m_N$.

In view of experiments, which are much easier to perform, we also
 evaluated the
semi-exclusive processes $^3{\rm He} (\gamma, p)$
and $^3{\rm He} (\gamma,n)$.
The cross sections are given as

\begin{equation}
{ {d^{\, 3} \sigma} \over { d \Omega_1 \, d E_1 } }
\ = \
{ 2 \pi^2 \alpha \over E_\gamma } \,
m_N^2 \,
{ 1 \over 2 } \, \mid {\vec p}_1 \mid \, \mid {\vec p} \mid \
{\cal C} \ \int d {\hat p} \,
{ 1 \over 2 } \,
\sum_{M,m_1,m_2,m_3} \, \left( \mid N_{+1} \mid ^2 + \mid N_{-1}
\mid ^2 \right) ,
\label{eq:sigma3}
\end{equation}
where $ \mid {\vec p} \mid $ and ${\hat p}$ are the magnitude
(kinematically fixed)
and the direction of the relative
momentum between nucleons 2 and 3.
${\cal C} = \frac12$ if the two unobserved particles
are identical and ${\cal C} = 1$ otherwise.

\section{Results}
\label{secIII}

Because of the lack of a full theoretical understanding of nuclear forces, a
possible way to search for 3NF effects is to use all present-day high
precision NN forces in 3N calculations and look for differences between
theoretical predictions and data for 3N observables. Such searches
have already been performed for the
binding energies of $^3$He, $^3$H and $^4$He~\cite{Nogga.2000},
cross sections and
spin observables in
elastic nucleon-deuteron
scattering~\cite{ref.spinstructure,Sekiguchi}
and in the nucleon induced deuteron breakup
process~\cite{ref.Report,Kuros1,Kuros2}. The inclusion of
various present-day 3NF models sometimes removes
the differences but sometimes do not.
Thus right now the properties of 3NF's are still not known.
In such a situation all possible information should be used
and
$^3$He photodisintegration is a good additional test ground to search
for 3NF effects.

We use various dynamical inputs: the high precision NN potential
AV18 \cite{ref.AV18}
together with the Urbana~IX 3NF~\cite{ref.urbanaIX}.
That model correctly describes the $^3$H
binding energy. In our calculations we neglect the pp Coulomb force in the 3N
continuum  but keep it in the $^3$He bound state. Except for 21 keV the
binding energy is then the correct one.
In addition to the standard nonrelativistic single nucleon current
operator, we employ explicit $\pi-$ and $\rho-$ like MEC's~\cite{Kotlyar}
according to the Riska
prescription~\cite{Riska}. They are consistent with dominant parts
of AV18 and fulfill the continuity equation in
relation to those parts. This has to be improved in the future by adding
the remaining pieces to be fully consistent to AV18 in the continuity
equation. As an alternative method,
we also use the Siegert theorem in the
form given in \cite{Golak.pdc.2000}. Also here
improvements are needed in the future to add explicit MEC's to the
magnetic multipoles which are not affected by the Siegert approach.
Therefore both approaches to many body currents leave room for
improvement.
That form does not use long wave length
approximations and is formulated in momentum space.
In order to have a first test of the dependence on the choice among the
various possible NN and 3NF combinations, we also used the NN force
CD Bonn ~\cite{CDBONN}
plus the modified Tuscon Melbourne TM' 3NF~\cite{ref.modTM}.
That modified force removes deficiencies of the older TM 3NF~\cite{ref.TM},
which was in conflict
with chiral symmetry~\cite{ctobe}.
In fact we use the newest set of parameters for this force, as given in~\cite{newNogga}.

 In order to search for 3NF effects in the 3N $^3$He photodisintegration,
we performed the following investigation.
We scanned the whole 3N phase space and compared the exclusive breakup cross
section based on NN forces only to the one adding 3NF's.
To that aim
we define the quantity

\begin{equation}
\Delta(\Omega_1, \Omega_2, S) \ \equiv \
\mid   {d^{\, 5} \sigma}^{\rm NN+3NF} - {d^{\, 5} \sigma}^{\rm NN} \mid  /
    {d^{\, 5} \sigma}^{\rm NN} \times 100\%\;.
\label{Delta}
\end{equation}
In this manner we can associate $\Delta$-values to all regions in
phase space. Such a
search is carried through using
two different NN and
3NF combinations: AV18 alone and combined with Urbana~IX, and CD Bonn alone and together
with TM'. Further, in the case of
AV18, we work either with
the MEC's explicitly or the Siegert approach combined with the single nucleon
current operator.
In the case of CD Bonn only the Siegert approach is chosen
 since consistent
MEC's are not available (in any case they would not be well defined
since that NN force has
been introduced partial wave per partial wave). In order to locate
 phase space regions
uniquely, we show three two-dimensional plots.
The first one is the $\Theta_1-\Theta_2$
plane for
the two angles of the proton detectors. The second one is the
$\Theta_1-\Phi_{12}$ plane,
where $\Phi_{12} \equiv \mid \Phi_1 - \Phi_2 \mid$ is the
relative azymuthal angle for the two detectors. Finally, the third one
is the E$_1$-E$_2$ plane
for the correlated energies of the two detected protons.
To fill the three planes we proceed as follows. The whole phase-space
is filled with discrete points corresponding to certain grids
in $\Theta_1,\Theta_2,\Phi_1,\Phi_2$, and E$_1$. For $\Theta_1$ and $\Theta_2$
fixed we search for the maximal value of $\Delta$ in the 3-dimensional
subspace spanned by $\Phi_1,\Phi_2$, and E$_1$. Then we combine
those maximal $\Delta$-values into three groups and associate certain
grey tones to those group values.
Next we choose a fixed $\Theta_1$ and $\Phi_{12}=\mid \Phi_2 \mid$ (one can put
$\Phi_1=0^{\circ}$) and search again for the maximal values of $\Delta$
in the 2-dimensional subspace spanned by  $\Theta_2$ and E$_1$.
The same grey tones and groupings are then applied. Finally, in the
E$_1$-E$_2$ plane we search for the maximal $\Delta$-values in the three
dimensional subspace spanned by $\Theta_1,\Theta_2,\Phi_{12}$ and repeat the
procedure. For a larger number of groups see~\cite{romek.thesis}.
This procedure will be now applied in Figs.\ref{fig1}-\ref{fig5}.
We performed the
investigation for three photon LAB energies E$_\gamma$= 12, 40 and 120 MeV.
Please note that in~\cite{romek.thesis} the NN interaction was taken in the
form of $np$-interaction only, while in present work we include
$pp$ and $nn$ interaction by the "$\frac{2}{3}+\frac{1}{3}$" rule~\cite{witala39}.

Fig.~\ref{fig1} is based on AV18+Urbana~IX and the use of the Siegert's approach.
For the sake of visibility, since
we use only grey tones, we split the variations of the quantity $\Delta$
into two groups which are explicitly shown and a third one in between
which is just white.
This is done for
each energy. A more refined splitting (shown in color) can be found
 in~\cite{romek.thesis}.
The first, second and third row refers to E$_\gamma$=12, 40 and 120 MeV,
 respectively.
Based on the meaning of the grey tones, as explained above, one can proceed as follows.
Choosing a region in the $\Theta_1-\Theta_2$ plane with a black tone we know
that in the $\Theta_1-\Phi_{12}$ plane there must exist also black region for the
same $\Theta_1$. This allows to read off a certain value of $\Phi_{12}$.
Then the angular positions of the two detectors are fixed, which defines the
S-curve in the E$_1$-E$_2$ plane. Along such a S-curve there must be again a black
region, where one can read off the corresponding range of energies.
Choosing for instance another combination of tones, like a black one in the
$\Theta_1-\Theta_2$ plane, white one in the $\Theta_1-\Phi_{12}$ plane one knows
that the S-curve in the E$_1$-E$_2$ plane lies in the white and maybe grey
regions. This should explain the use of all Figs.\ref{fig1}-\ref{fig5}.
Clearly, the biggest 3NF effects are for E$_\gamma$=120 MeV reaching up
 to
85\%. Thus
for instance for angular configurations $\Theta_1=\Theta_2 \approx $40$^\circ$,
 $\Phi_{12} \approx 20^{\circ}$
and for instance E$_1 \approx $20 MeV and E$_2 \approx $20 MeV 3NF
effects of that big size occur
for that nuclear force model and for that choice of the  electromagnetic
 current operator. At E$_\gamma$=40 MeV the
effects are significantly smaller, which comes us a surprise since they are
larger again at 12 MeV. Maybe
it is a similar phenomenon as we found  in~\cite{Romek.2b} for the pd breakup process
in $^3$He photodisintegration. There  we saw that 3NF effects essentially
 vanished around
30 MeV, whereas below and above that energy they were significantly
 present.
Here, at the lowest energy E$_\gamma$=12 MeV
3NF effects are as large as
50\%. The white areas between
 the dark and grey
shaded regions in the two left panels for all three energies
refer to $\Delta$ values
between 20 and 30\% in case of E$_\gamma$=12 MeV and correspondingly for the other
energies. In
the very right panels the allowed energies E$_1$ and E$_2$ are kinematically restricted
and events between
20 and 30\% for E$_\gamma$=12 MeV etc. are present between the dark and grey shaded regions,
whereas in the right upper corner there are no events.

This result can now be compared to the choice CD Bonn+TM' in Fig.\ref{fig2}, again
 using the Siegert
approach.
For E$_\gamma$=12 and 120 MeV the outcome is qualitatively similar
to Fig.\ref{fig1}, except that
for 120 MeV the dark spots around $\Theta_1=\Theta_2 \approx $40$^{\circ},
\Phi_{12}\approx$20$^{\circ}$
and E$_1$=E$_2 \approx $20 MeV
are missing. At 40 MeV, where the effects are small,
the patterns are nevertheless in reasonable agreement (for a more detailed comparison
see~\cite{romek.thesis}).

 Now we ask the question, will the choice of handling MEC's disturb the outcome too strongly?
To give a first hint
to the answer, we show in Fig.\ref{fig3} the choice AV18+Urbana~IX now together
 with explicit MEC's instead of Siegert.
Comparing to Fig.\ref{fig1}, the patterns are at least qualitatively
 similar. This is desirable, since both current prescriptions should be close to
 each other, after all.
But there are differences which in a quantitative analysis of future data
might be disturbing.
This has been quantified by comparing the cross sections
underlying Figs.\ref{fig1} and~\ref{fig3} and
locating the phase- space regions where that difference is
 large or small. We find that at 12 MeV the difference in the two approaches for
the currents stays below about 20\% in most of the phase space regions, whereas
already at 40 MeV it is roughly only in half of the phase space region.
At 120 MeV the difference is larger. Clearly the question of the choice of the
current requires further theoretical investigations which, however, is outside the
scope of this paper.
We refer the reader to~\cite{romek.thesis} for more details.

Let us now add two comments.
It is important to note that a single nucleon current operator alone
 would be totally
insufficient. This is a well known fact for photodisintegration. We
 demonstrate this by
defining

\begin{equation}
{\Delta}'(\Omega_1, \Omega_2, S)  \ \equiv \
\mid   {d^{\, 5} \sigma}^{\rm NN+3NF}_{MEC} -
{d^{\, 5} \sigma}^{\rm NN+3NF}_{\rm single \ nucleon} \mid /
    {d^{\, 5} \sigma}^{\rm NN+3NF}_{\rm single \ nucleon} \times 100\%
\label{Deltatwoprime}
\end{equation}
and display the corresponding regions in phase space in Fig.\ref{fig5}.
The force combination AV18+Urbana~IX has been used.
The outcome is clear cut. In most regions of the phase space $\Delta'$
 is much larger than
100\%
at higher energies.
Even at 12 MeV there are many phase space regions, where
 using the single nucleon
current operator would be wrong by about
50\%.

Often in the literature photodisintegration is treated keeping only the
 lowest multipole
E1. This extreme low energy assumption would be quite insufficient
for nearly all phase
space regions and for all three photon energies studied in this paper.
This can again be quantified and we find, that even at
 12 MeV there are plenty of breakup
configurations where the electric multipole E1 alone would be wrong by more than 20\%.
Again for  detailed plots see~\cite{romek.thesis}.

Finally, but quite important for future experiments, we display the
regions in phase space
where    for AV18+Urbana~IX based on the Siegert approach
and all multipoles (in practice up to E7 and M7)
$ {d^{\, 5} \sigma}^{\rm NN+3NF}$
takes
on certain values. This
is shown in Fig.\ref{fig7}.
We divided the cross section values for each photon energy arbitrarily
 into three groups. Now the white regions contain cross section values
below the lowest values explicitly stated.
In the $\Theta_1-\Theta_2$ planes the prominent enhancements are along
$\Theta_1 \approx \Theta_2$ and
$\Theta_2  \approx  180^{\circ}-\Theta_1$. They are connected to proton-neutron
and proton-proton final state interaction peaks (FSIP)
configurations, respectively. The pp FSIP's occur for small $\Phi_{12}$'s and the
pn ones for
larger $\Phi_{12}$'s. The FSIP character is clearly documented in the E$_1$-E$_2$
projections with a
high energy transfer to one of the nucleons (in  case of both E$_1$ and E$_2$ low the high
energy transfer is of course to the neutron).

We would like to point to regions in phase space,
where the cross
section is large, 3NF effects are large and the difference
 in the predictions choosing
Siegert or explicit MEC's is small. For a certain
 quantification of those requirements we
display the results in Fig.\ref{figZB}. This should be of special interest
for future experiments. In Fig.\ref{figZB} all three rows are for
E$_\gamma$=120 MeV.

 Now we would like to show a few examples for the five fold differential
 cross sections
directly.
First, we regard a case corresponding to Fig.\ref{figZB},
 where 3NF effects are large,
the difference between the current predictions is small
 and the cross section is large.
This is shown in Fig.\ref{konf51}. Another extreme and opposite
 case is displayed
 in Fig.\ref{konf52}, where the two
choices of currents lead to large differences but where
 3NF effects are small. In Fig.\ref{konf53}
we show a case where more than one-body current effects
 are especially large and finally
in Fig.\ref{konf56} a case where all possible force  and current combinations
 give essentially the same
result. Thus we see a great variety in the interplay
 of forces and currents for different
asymptotic configurations.
We refer to~\cite{romek.thesis} for results for E$_\gamma$=12 and 40 MeV and
where moreover in addition to $^3$He also the $^3$H target has
 been considered.

 Before we compare to a few existing data we would like to show the semi-exclusive cross
sections for $^3$He and ejecting either a proton or a neutron. This is displayed in
Figs.\ref{figSE1}-\ref{figSE2}
for the example E$_\gamma$=120 MeV and  for four  selected ejection angles. For the other
energies, 12 and 40 MeV, see~\cite{romek.thesis}.
We show the following force and current combinations:
AV18+single nucleon current, AV18+Siegert, AV18+MEC, AV18+Urbana IX + Siegert. The 3NF
effects are unfortunately rather small. The integration over the two spectator nucleons
(the two angles of their relative momenta) for each given nucleon energy E$_{p,n}$
washes out the
strong signatures for 3NF's which are located only in part of the integrated phase space as
seen in Fig.\ref{fig1}.
Nevertheless data for this relatively "simple" one-arm experiment would be
of interest to test theory in that partially integrated form as we demonstrate now.

 For the p- ejection we see four peak structures, two of them of the type FSIP.
The relative  energy of two outgoing nucleons goes to zero and one sees
the enhancement  of the NN t-matrix due to the virtual $^1$S$_0$ state. The structure for
instance for $\Theta_p$=60$^\circ$ at E$_p$=20 MeV is a pp FSIP, which is shifted
slightly for the other proton angles. In case of the neutron ejection the corresponding
peaks are due to a pn pair. The FSIP's at the highest nucleon energies are due to  the
not detected  pairs, pn for p- ejection and pp for n- ejection.
The pronounced peak around 74
MeV for $\Theta_n=0^{\circ}$ and the corresponding shifted ones for the other neutron  angles
are due to a complex interplay of the phase space factor, enhancement in the $^3$He wave
function due to certain momentum arguments, final state interactions and two-body
currents. To achieve this sort of insight, we looked first into PWIAS\footnote{PWIAS
denotes the fully antisymmetrized plane wave approximation} alone and using only
the single nucleon current. In that case which allows analytical insight, we found that the
enhancement results from small momentum arguments in the $^3$He wave function (about 1
fm$^{-1}$).
(The momentum dependence of a 3N wave function is nicely displayed for instance
in~\cite{elster}).
That peak structure in PWIAS survives if one adds the other dynamical ingredients.
Thereby we investigated under the full dynamics the individual kinematically complete
contributions of the five fold  differential cross section to the semi exclusive one in that
peak
region. We found that the dominant contributions arise from nearly back to back breakup
configurations. The neutron is ejected under 0$^{\circ}$, to choose one example, and one
proton close to 160$^{\circ}$ with an energy of 34 MeV.
 The second proton
receives very little energy (a few MeV). In case of the proton ejection the
corresponding peak around 78 MeV and $\Theta _p$ = 0$^{\circ}$ receives again the dominant
contribution from the proton in forward direction and  a neutron in backward direction
with energies as above.
Again  the second proton has a very small energy. The peak at the  very low
proton  energy spectrum gets its dominant contribution from a proton-neutron pair emitted
 roughly
back to back and about perpendicular to the photon direction. Each one has
about 50 MeV. That peak structure, which is also seen in PWIAS, is absent
in case of the n- ejection.
We could not clarify that point satisfactorily. But we found that if we remove the
channels in case of proton-ejection where the "spectator pair" of a proton and a neutron
interacts in the states $^3$S$_1$-$^3$D$_1$ the peak is dramatically reduced.
Such a state is absent for the pp-pair in case of the neutron ejection. It is
also of interest to point to the fact that those peaks with the underlying structure of
back- to- back emission are strongly enhanced by the action of two-body currents. That
enhancement is  much reduced outside those peak regions. Because of that interesting
underlying dynamics comparison to data would be very welcome.

We do not show the corresponding curves for the semi-exclusive process for $^3$H
since they are very similar
 in shape if
proton and neutron are replaced against each other~\cite{romek.thesis}.

Now let us finally come to data.
As mentioned in the Introduction there are data~\cite{ref1} for
$^3{\rm He}(\gamma, pp)n$  for photon energies
between E$_\gamma$=90 and 250 MeV.
Table I of~\cite{ref1} shows the central proton detection angles for the 4 angular
combinations chosen in that experiment. For fixed angles of the two
proton detectors the proton energies are correlated and kinematically
allowed events have to lie on a locus, as pointed out before.
This corresponds to the
representation of the five fold differential cross section we used before
in Figs.~\ref{konf51}-\ref{konf56}. We show in Figs.\ref{figSA1} and~\ref{figSA2}
the cross section $d^5\sigma / d\Omega_1 d\Omega_2 dS$ for two
examples (LR-RL and LL-RR configurations, using the notation of~\cite{ref1}) from the four
angular combinations studied experimentally in~\cite{ref1}.
Among the four photon energies we looked into
(E$_\gamma$=80, 100, 120 and 160 MeV), only the highest
is above the pion threshold.
We compare cross
sections for AV18 and AV18+Urbana~IX. In all cases the explicit MEC's have
been used. While for the first angular combination 3NF effects are hardly
visible they can be seen for the second angular combination though the
effects stay below 25\%. The cross sections for the two remaining
angular combinations from~\cite{ref1} (not shown) are somewhere in between
and the 3NF effects are rather small.
In principle these results should
be compared to the data. The data of~\cite{ref1} were, however, integrated over
the S-curve (see below).

We compare our theory to those integrated cross sections.
We took the cuts for the minimal proton energies quoted
in~\cite{ref1} into account, but no further angular averaging.
These data are differential cross sections
 in both solid angles: $d^4\sigma / d\Omega_1 d\Omega_2 $.
Fig.\ref{figSA3} confronts our theoretical results based on AV18 +MEC and AV18+Urbana~IX
+MEC to the experimental data from Fig.~9 in~\cite{ref1}.
First of all we see that the 3NF effects are smaller than the error
bars and,  secondly, we can only state that we predict the right order of
magnitude. The reasonable agreement in case of the first angular
combination might be accidental. Certainly new and precise data would be
very welcome and a theoretical analysis,
which takes into account all experimental conditions,
should be carried through.
We note that we predict much larger 3NF effects for angular combinations
according to Fig.~\ref{figZB} (in case of E$_\gamma$=120 MeV).

The second experiment~\cite{ref2} quoted in the introduction shows
$^3$He($\gamma$,np)
differential cross sections as a function of the opening angle between the
neutron and the proton for $\Theta_p$=81$^\circ$ in the laboratory frame. We
compare our results for two photon energies E$_\gamma$=55 and 80 MeV to this
data. No angular averaging whatsoever has been performed in the theory. If we
look into Fig.~\ref{figFK1} we see again small 3NF effects and find a reasonable agreement
with the data except that our peaks are too high. At least partially this
might be related to the missing angular averaging and possible further
experimental conditions, which we could not take into account.

Finally, in Fig.\ref{figFT}, we compare theory for the total $^3$He and $^3$H 3N
photodisintegration
cross section to data in the low energy region. Some of the results have been
shown before in~\cite{Trento}. In both cases theory is roughly inside the bulk
of the data and in case of $^3$H the 3NF effects seem to be favored by the
data. Unfortunately, the quality of the data is not too high and precise data at low and
higher energies  are badly missing.

\section{Summary}

We performed Faddeev calculations for the $^3$He photodisintegration into three nucleons. The
NN forces AV18 and CD-Bonn in combination with the 3NF's Urbana IX and TM'
have been applied.
Results are presented for photon  energies E$_\gamma$=12, 40, and 120 MeV as representative
examples. We scanned the whole phase space for 3N breakup to search for regions,
where 3NF effects show up significantly. We found effects as large as
85\% which
should be checked experimentally. We also found that two-body currents are extremely
important and the restriction to  a single nucleon current would be rather meaningless. We
use explicit $\pi$- and $\rho$-like exchange currents consistent to the NN force AV18 and also
parallel to that the Siegert approach without long wavelength approximation. Both currents
lead qualitatively to the same results but not quantitatively, which clearly calls for an
improved future treatment.
Precise future data for that complete $^3$He breakup preferably for  all of the phase space
would be very useful to check the present day nuclear dynamics and the choice of the
electromagnetic current operator. Those data would supplement the search for 3NF
effects going on in 3N scattering~\cite{ref.spinstructure}-\cite{Kuros2}.

In addition, we predicted cross sections for the semi-exclusive processes $^3$He($\gamma$,p)
and $^3$He($\gamma$,n), where interesting peak structures occur in the energy dependence of the
knocked out nucleon.

Finally, we compared theory to data for the exclusive $^3$He breakup process. The comparison
was unfortunately hindered by the fact, that due to the lack of information,
we were not able
 to take the experimental conditions (acceptances in energy and angular resolutions etc.)
into account. Nevertheless, the at least qualitative agreement with the data shows that a
proper analysis of new data would be very valuable to find out how well theory describes the
complex interplay of NN and 3NF's with the absorption mechanism of the photon.

 The comparison with the total $^3$He breakup data was also inconclusive because the
available data below 30 MeV have large error bars and do not agree with each other.
Also data above 30 MeV are needed.

\acknowledgements
This work was supported by
the Deutsche Forschungsgemeinschaft (J.G.,R.S.),
the Polish Committee for Scientific Research
under Grants No. 2P03B02818
and 2P03B05622,
and by NFS grant No. PHY0070858.
R.S. acknowledges financial support of the Foundation for Polish Science.
The numerical calculations have been performed
on the Cray T90, SV1 and T3E of the NIC in J\"ulich, Germany.

\begin{figure}[h!]
\caption[ ]
{
The regions in the 3N phase-space projected onto the
$\Theta_1-\Theta_2$, $\Theta_1-\Phi_{12}$ and E$_1$-E$_2$ planes
carrying certain values of the quantity $\Delta$ from Eq.(\ref{Delta})
as indicated in the boxes.
That quantity is a measure for 3NF effects in
${ {d^{\, 5} \sigma} \over { d \Omega_1 \, d \Omega_2 \, d S } }$
and varies for
the three photon LAB energies shown in the first row (12 MeV), the second
row (40 MeV) and the third row (120 MeV). The force combination
AV18+Urbana~IX together with the Siegert approach has been used.
}
\label{fig1}
\end{figure}

\newpage

\begin{figure}[h!]
\caption[ ]
{
The same as in Fig.\ref{fig1} for the force combination CD Bonn+TM'
and the Siegert approach.
}
\label{fig2}
\end{figure}

\newpage

\begin{figure}[h!]
\caption[ ]
{
The same as in Fig.\ref{fig1} but instead of the Siegert approach
explicit MEC's have been used.
}
\label{fig3}
\end{figure}

\newpage

\begin{figure}[h!]
\caption[ ]
{
The same projections of the 3N phase space as in Fig.\ref{fig1} for
the quantity ${\Delta}'$ of Eq.(\ref{Deltatwoprime}).
${\Delta}'$ is a measure of the difference
in ${ {d^{\, 5} \sigma} \over { d \Omega_1 \, d \Omega_2 \, d S } }$
between the use of MEC's and the restriction of the
current operator to a single nucleon one. AV18+Urbana~IX has been
used.
}
\label{fig5}
\end{figure}

\newpage

\begin{figure}[h!]
\caption[ ]
{
The same projections as in Fig.\ref{fig1} for
${ {d^{\, 5} \sigma} \over { d \Omega_1 \, d \Omega_2 \, d S } }$
itself. AV18+Urbana~IX together with the Siegert approach has been used.
}
\label{fig7}
\end{figure}

\newpage

\begin{figure}[h!]
\caption[ ]
{
The same projections as in Fig.\ref{fig1} for
${ {d^{\, 5} \sigma} \over { d \Omega_1 \, d \Omega_2 \, d S } }$
itself for E$_\gamma$=120 MeV (all three rows).
AV18+Urbana~IX together with the Siegert approach has been used.
The additional conditions ($\sigma>\dots$) onto ${ {d^{\, 5} \sigma} \over { d \Omega_1 \, d \Omega_2 \, d S
} }$ in units of [$\mu$b sr$^{-2}$MeV$^{-1}$],
onto 3NF effects (3nf$>\dots$) and onto the difference between predictions
in the Siegert and explicit MEC
approaches (MEC$<\dots$),
are shown in the boxes for each line.
}
\label{figZB}
\end{figure}

\newpage

\begin{figure}[h!]
\leftline{\mbox{\epsfysize=200mm \epsffile{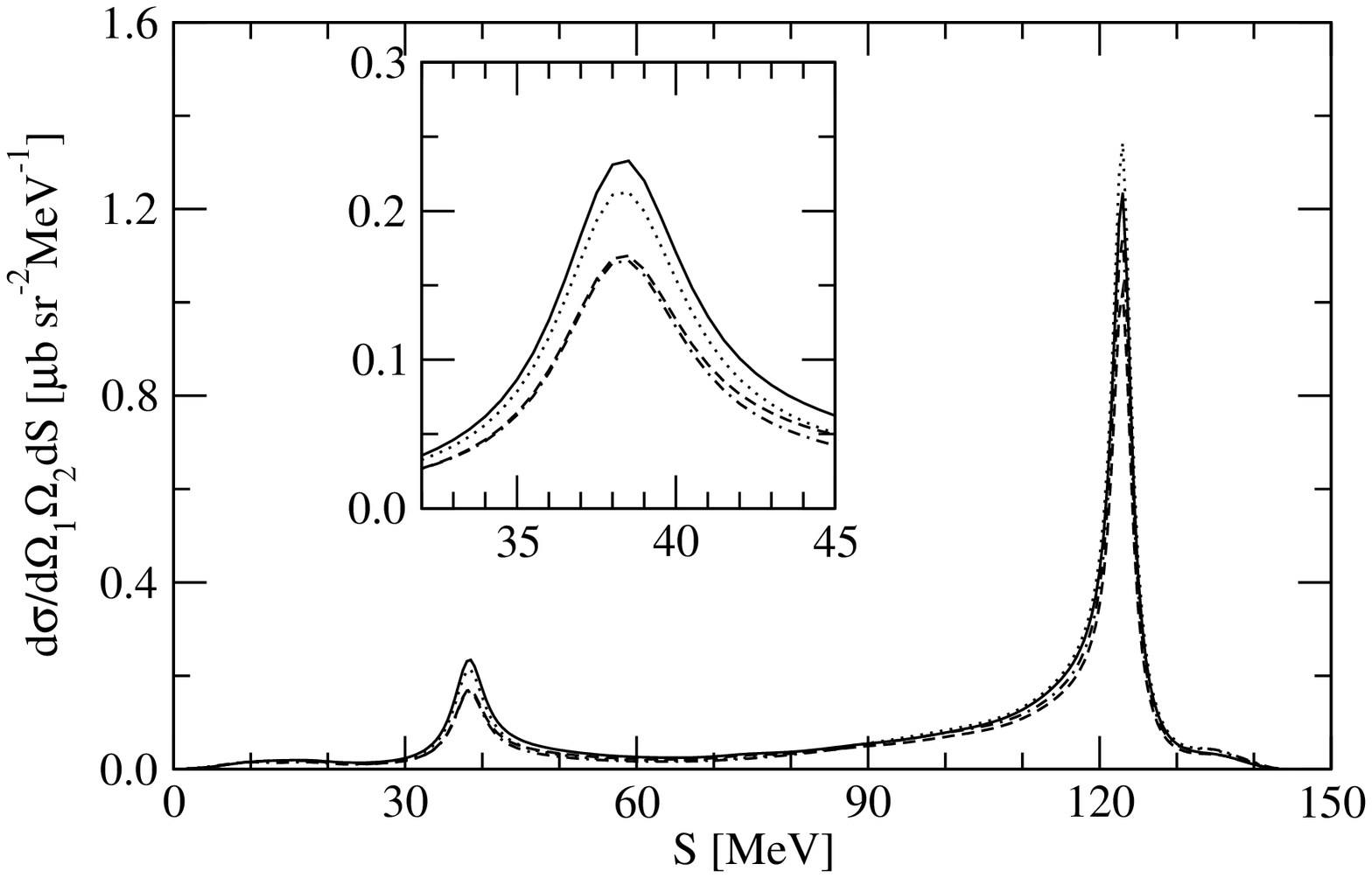}}}
\caption[ ]
{
Five-fold differential cross sections for the angular
con\-fi\-gu\-ra\-tion $\Theta_1=142^\circ$, $\Phi_1=0^\circ$,
$\Theta_2=27^\circ$, $\Phi_1=180^\circ$, at photon LAB energy
E$_\gamma$=120 MeV. The AV18 predictions in Siegert
approximation to the nuclear current (explicit MEC) are given by
dashed (dash-dotted) curve and corresponding AV18+Urbana~IX predictions
are given by solid (dotted) curve.
}
\label{konf51}
\end{figure}

\begin{figure}[h!]
\leftline{\mbox{\epsfysize=200mm \epsffile{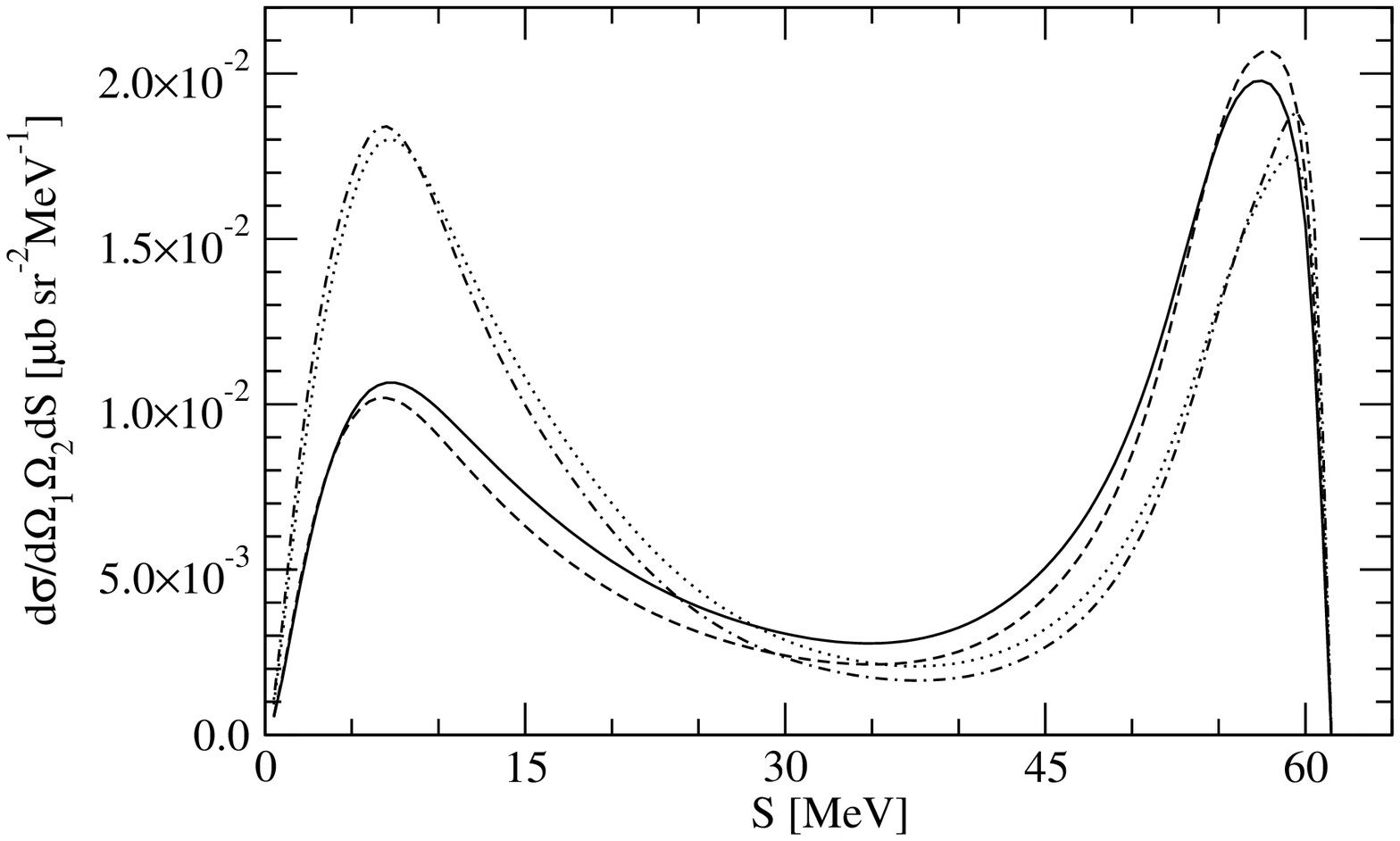}}}
\caption[ ]
{
The same as in Fig.\ref{konf51} but for
$\Theta_1=101^\circ, \Phi_1=0^\circ,
\Theta_2=164^\circ, \Phi_1=109^\circ$.
}
\label{konf52}
\end{figure}

\begin{figure}[h!]
\leftline{\mbox{\epsfysize=200mm \epsffile{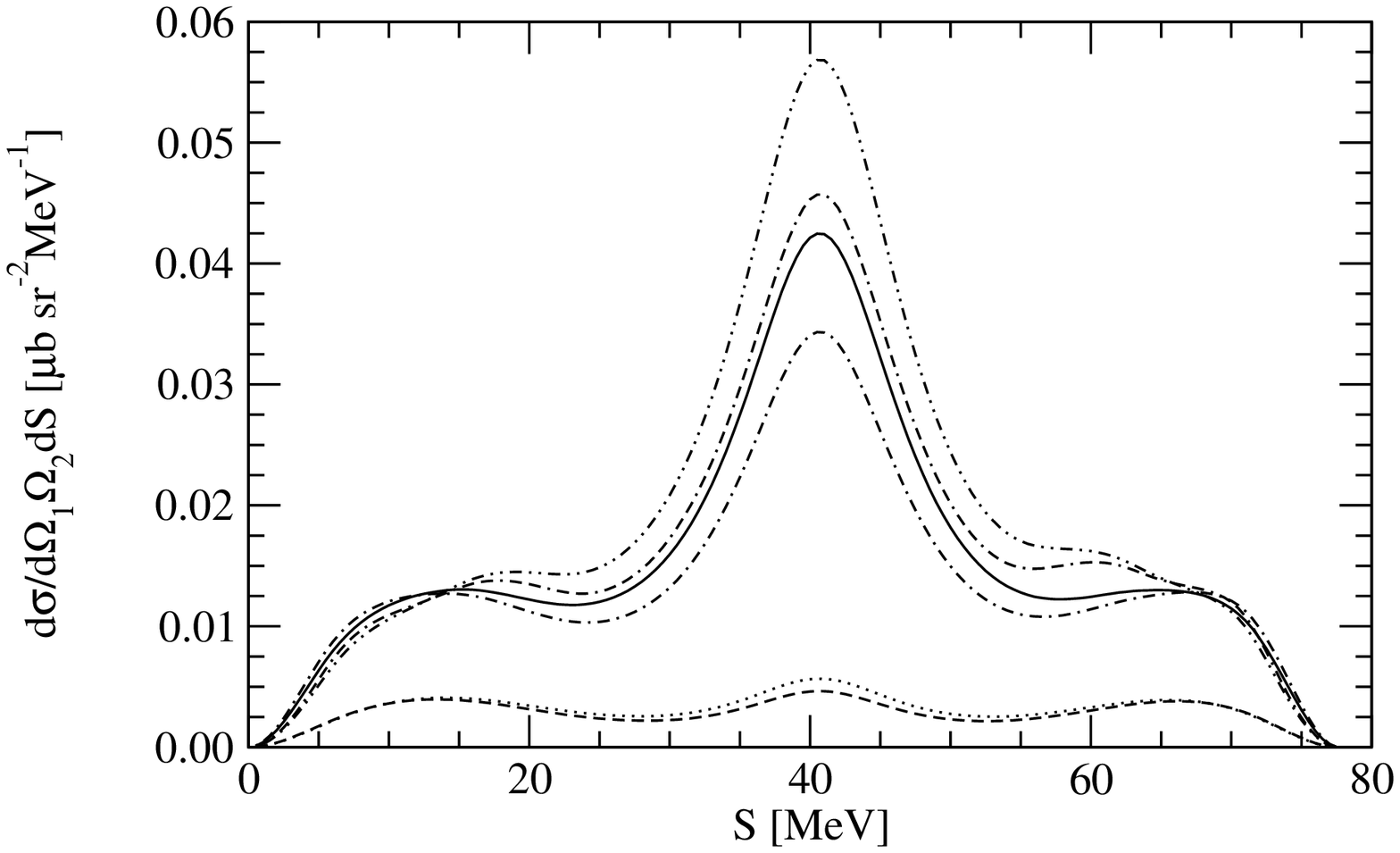}}}
\caption[ ]
{
Five-fold differential cross sections for the angular
con\-fi\-gu\-ra\-tion
$\Theta_1=88^\circ$, $\Phi_1=0^\circ$,
$\Theta_2=100^\circ$, $\Phi_1=11^\circ$, at
photon LAB energy E$_\gamma$=120 MeV. The AV18 predictions in single-nucleon,
Siegert and explicit MEC
approximations to the nuclear current are represented by
dashed, dotted-double-dashed and dash-dotted curves, respectively.
The corresponding AV18+Urbana~IX predictions
are represented by dotted, dashed-double-dotted and solid curves, respectively.
}
\label{konf53}
\end{figure}

\begin{figure}[h!]
\leftline{\mbox{\epsfysize=200mm \epsffile{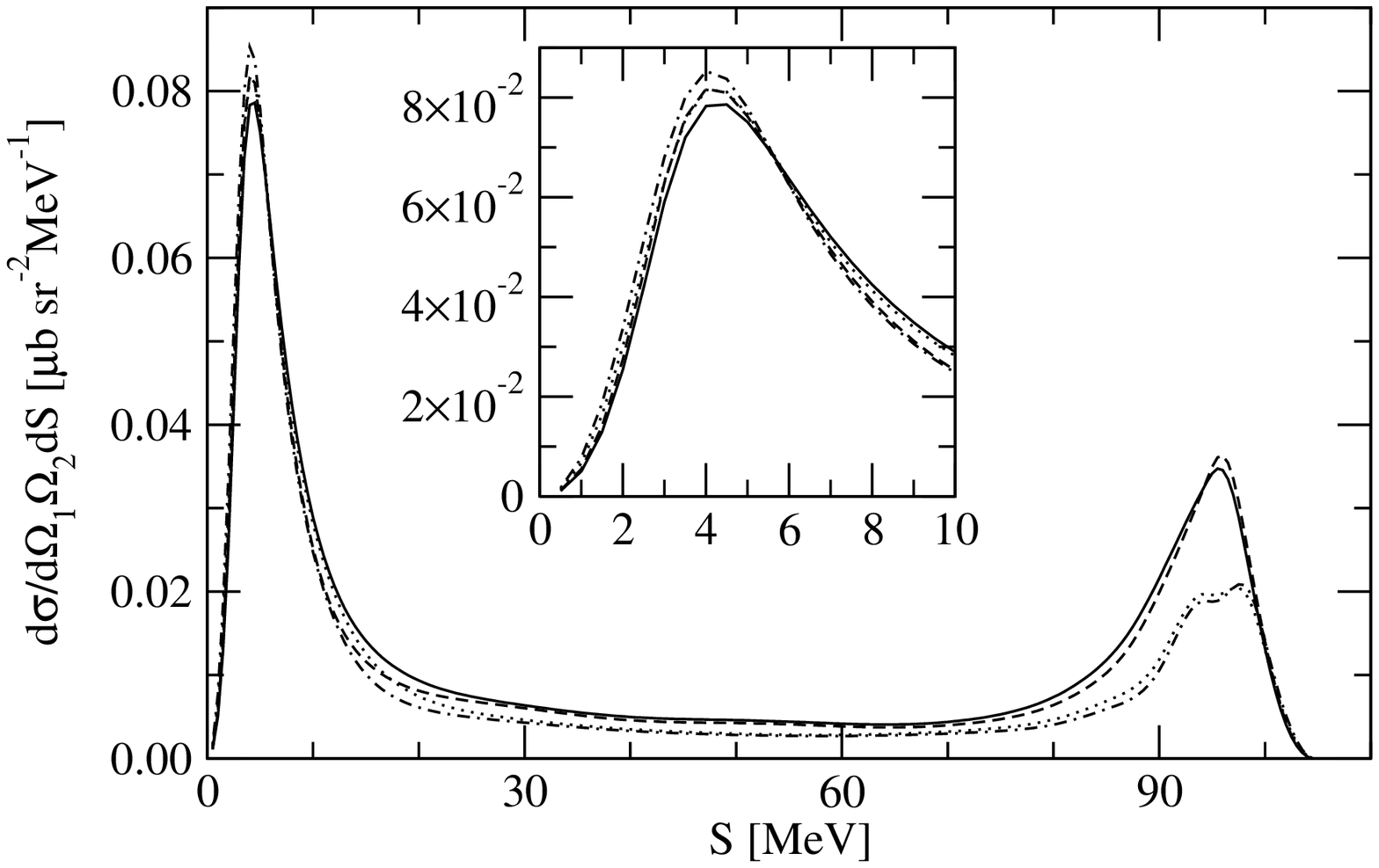}}}
\caption[ ]
{
The same as in Fig.\ref{konf51} but for
$\Theta_1=30^\circ, \Phi_1=0^\circ,
\Theta_2=145^\circ, \Phi_1=77^\circ$.
}
\label{konf56}
\end{figure}

\begin{figure}[h!]
\leftline{\mbox{\epsfysize=200mm \epsffile{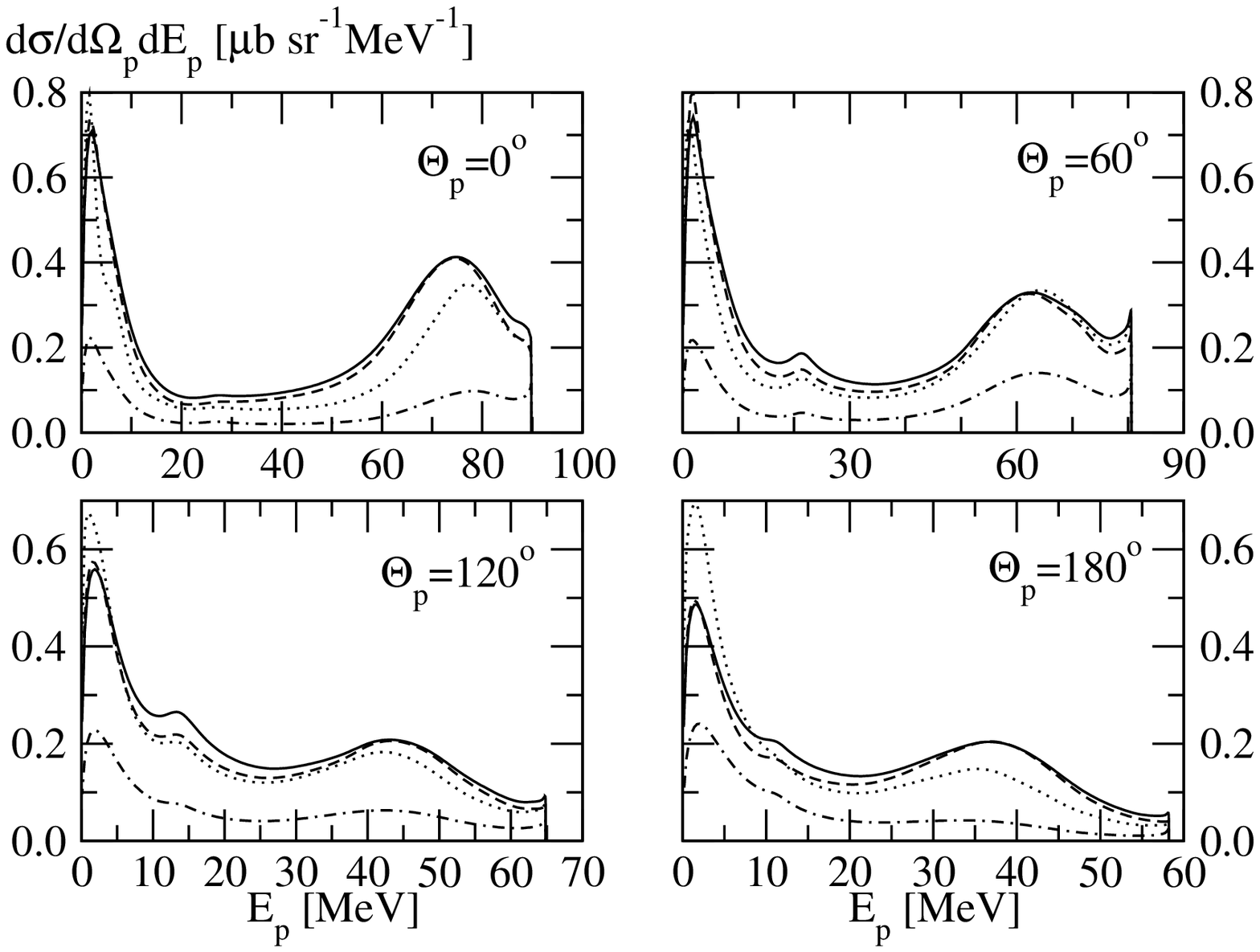}}}
\caption[ ]
{
The semi-exclusive cross sections for the process
$^3{\rm He} (\gamma, p)$
for four ejection angles
as a function of the proton energy E$_p$.
The solid curve is for AV18+Urbana~IX+Siegert,
the dashed curve for AV18+Siegert, the dotted curve for
AV18+MEC and the dashed-dotted curve for AV18+single nucleon
current operator. The photon energy is E$_\gamma$= 120 MeV.
}
\label{figSE1}
\end{figure}

\begin{figure}[h!]
\leftline{\mbox{\epsfysize=200mm \epsffile{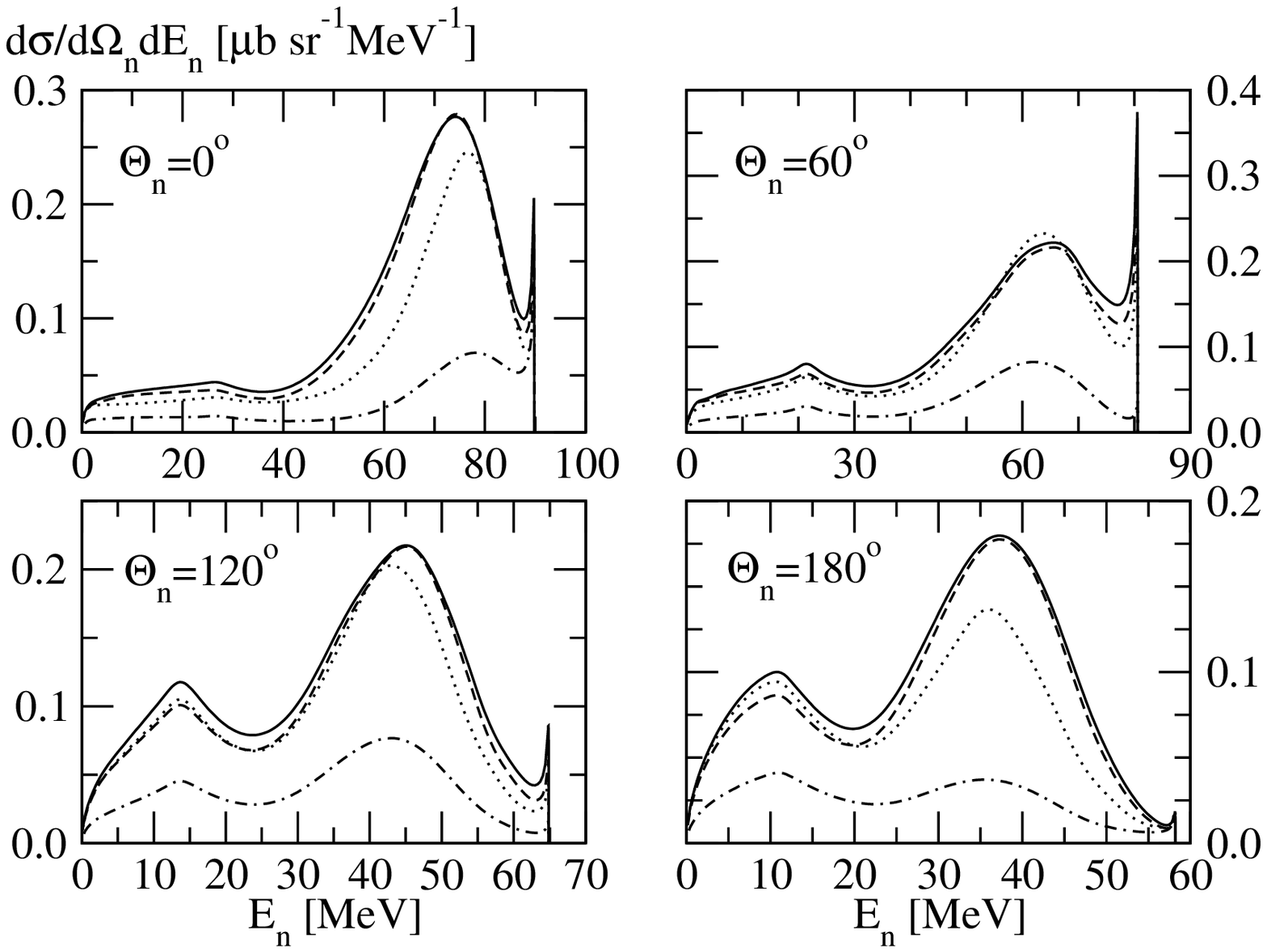}}}
\caption[ ]
{
The semi-exclusive cross sections for the process
$^3{\rm He} (\gamma, n)$
for four ejection angles
as a function of the neutron energy E$_n$.
Curves as in Fig.~\protect\ref{figSE1}
}
\label{figSE2}
\end{figure}

\begin{figure}[h!]
\leftline{\mbox{\epsfysize=180mm \epsffile{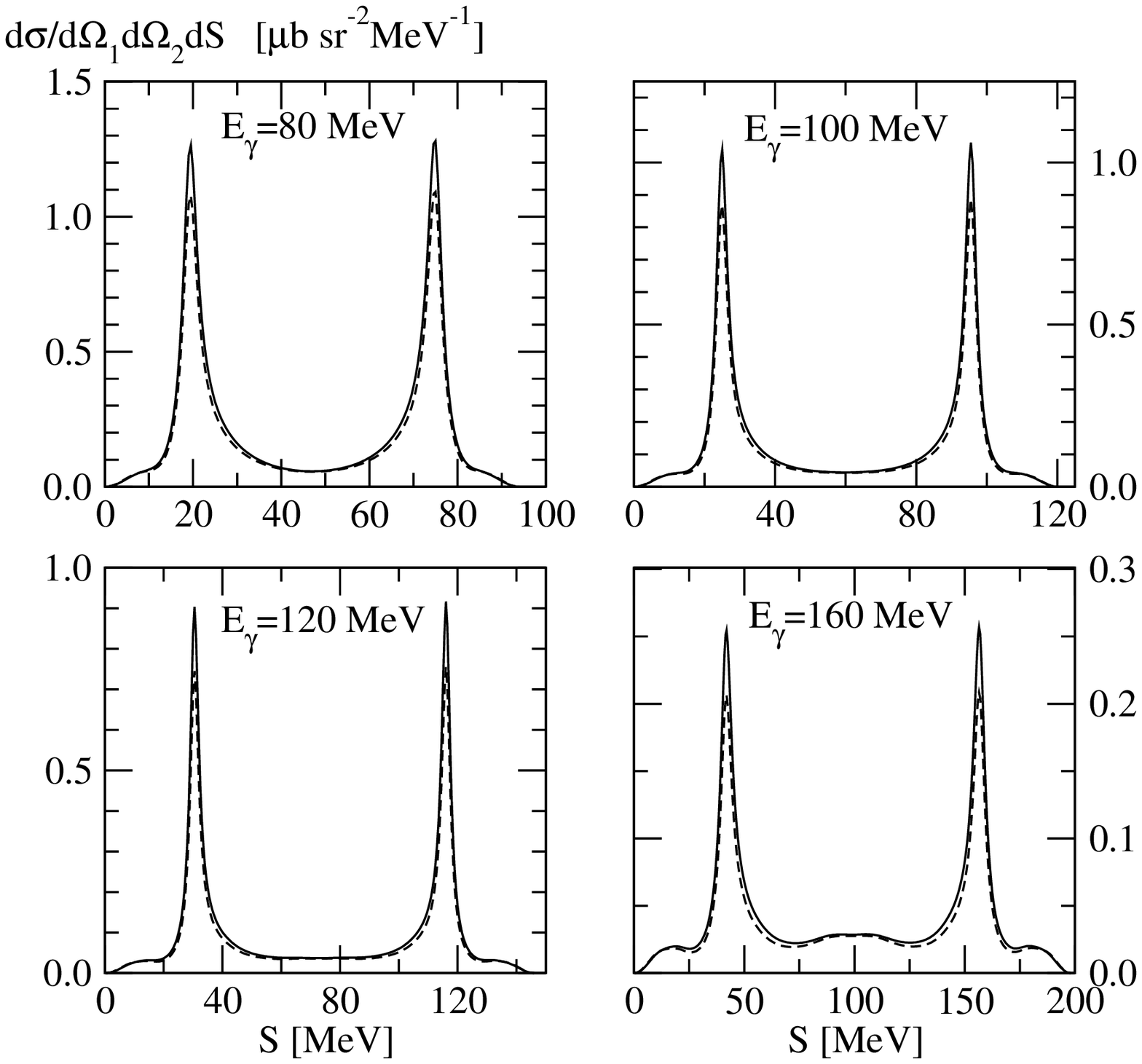}}}
\caption[ ]
{
Five-fold differential cross sections for the angular
configuration LR-RL of~\cite{ref1} $(\Theta_1=81.0^\circ, \Phi_1=0.0^\circ,
\Theta_2=80.3^\circ, \Phi_2=180.0^\circ)$ along the S-curve for the photon
energies $E_\gamma$= 80, 100, 120 and 160 MeV. The predictions for
AV18 (dashed curve) and AV18+Urbana~IX (solid curve) are
compared.
}
\label{figSA1}
\end{figure}

\begin{figure}[h!]
\leftline{\mbox{\epsfysize=180mm \epsffile{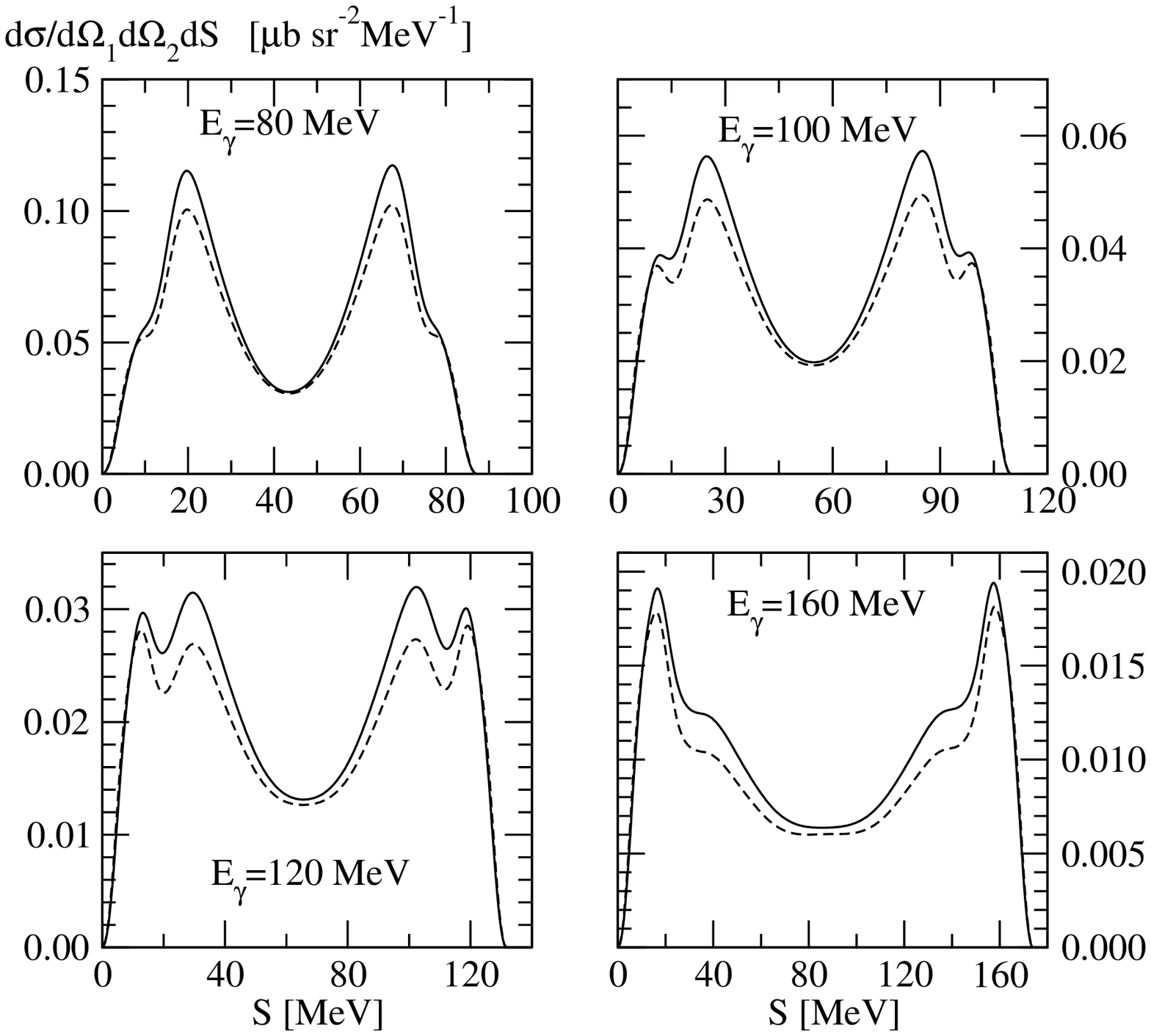}}}
\caption[ ]
{
The same as in Fig.~\protect\ref{figSA1} for the angular configuration
LL-RR of~\cite{ref1} $(\Theta_1=92.2^\circ, \Phi_1=0.0^\circ,
\Theta_2=91.4^\circ, \Phi_2=180.0^\circ)$.
}
\label{figSA2}
\end{figure}

\begin{figure}[h!]
\leftline{\mbox{\epsfysize=180mm \epsffile{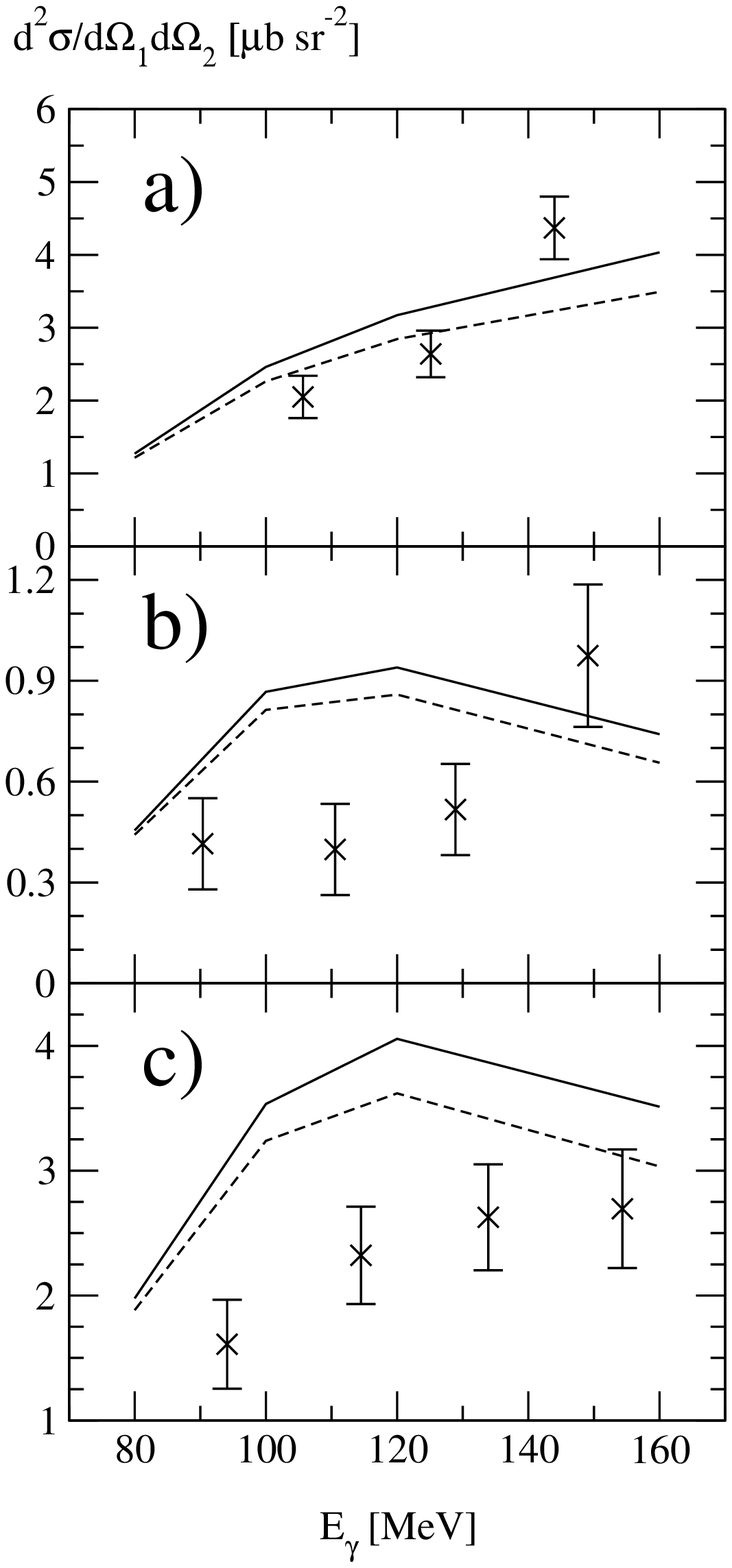}}}
\caption[ ]
{
The four-fold differential cross sections
${ {d^{\, 4} \sigma} \over { d \Omega_1 \, d \Omega_2 } }$
for the $^3$He($\gamma$,pp)n process
as a function of $E_\gamma$ in comparison to data given in
Fig.~9 of~\cite{ref1} for the angular configurations LR-RL (a) ,
LL-RR (b) and LL-RL+LR-RR
$(\Theta_1=91.7^\circ, \Phi_1=0.0^\circ,
\Theta_2=80.9^\circ, \Phi_2=180.0^\circ)$ and
$(\Theta_1=81.5^\circ, \Phi_1=0.0^\circ,
\Theta_2=90.8^\circ, \Phi_2=180.0^\circ)$
(c). The solid curve
is for AV18+Urbana~IX+MEC, the dashed curve for AV18+MEC.
}
\label{figSA3}
\end{figure}

\begin{figure}[h!]
\leftline{\mbox{\epsfysize=180mm \epsffile{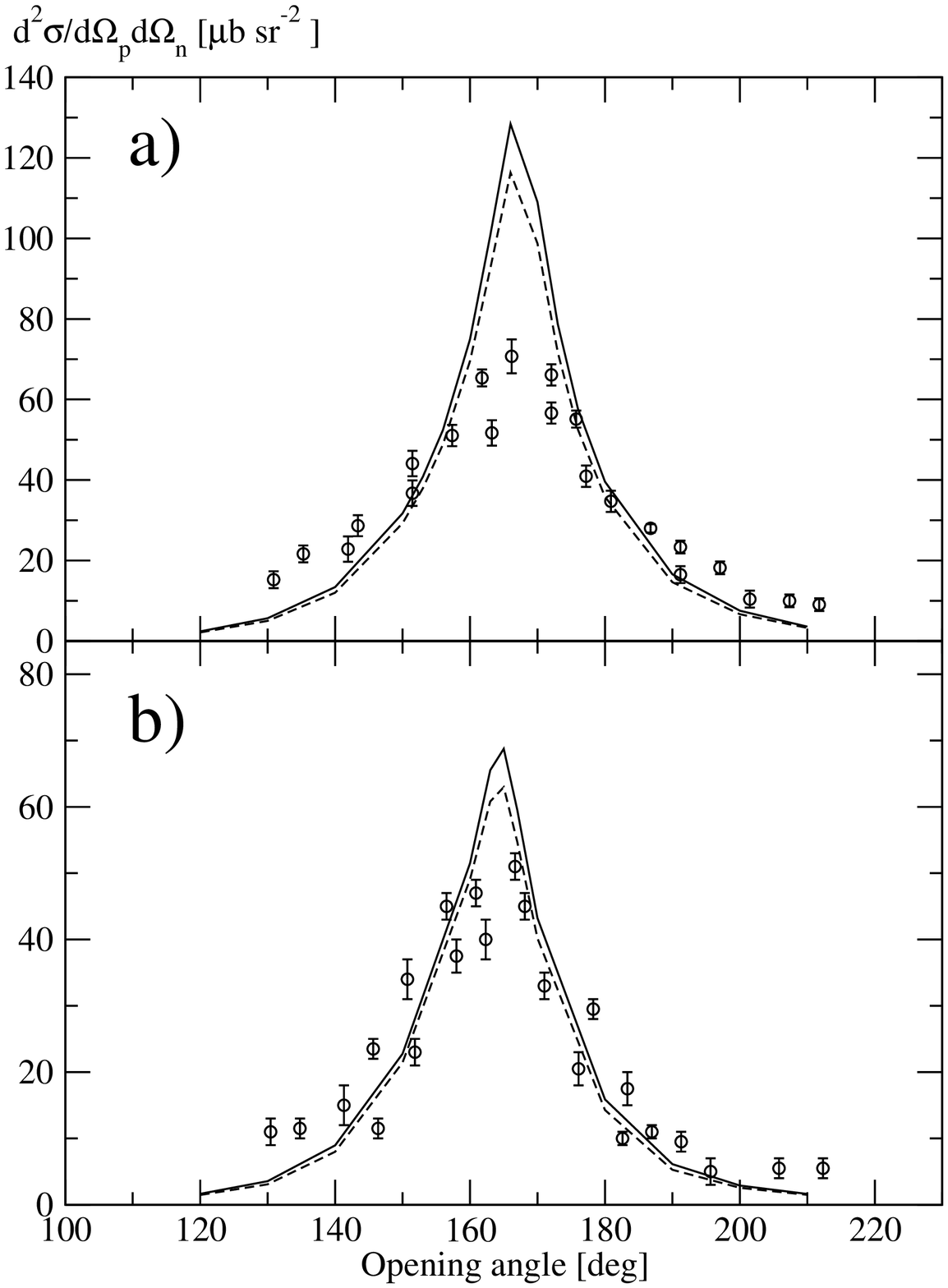}}}
\caption[ ]
{
The four-fold differential cross sections
${ {d^{\, 4} \sigma} \over { d \Omega_1 \, d \Omega_2 } }$
against the opening angle at $E_\gamma$= 55 (a) and 80 MeV (b)
for the $^3$He($\gamma$,pn)p process
in comparison to data from~\cite{ref2}. The data in (b) are
taken for $E_\gamma$= 85 MeV. Solid and dashed curves as in
Fig.~\protect\ref{figSA3}.
}
\label{figFK1}
\end{figure}

\begin{figure}[h!]
\leftline{\mbox{\epsfysize=180mm \epsffile{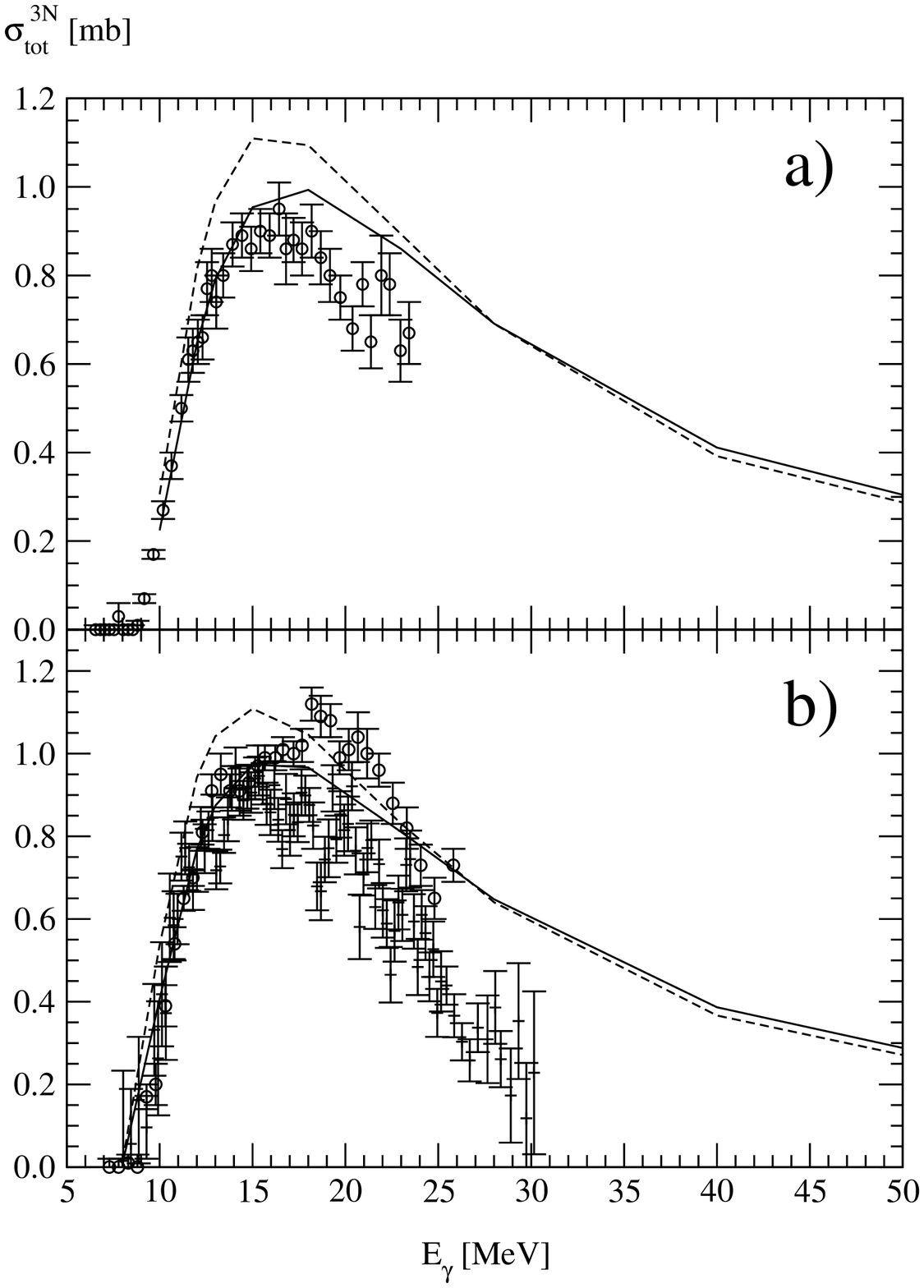}}}
\caption[ ]
{
The total breakup cross sections for $^3$H (a) and $^3$He (b)
photodisintegration in comparison to data. The solid curve refers
to AV18+Urbana IX, the dashed curve to AV18 alone, both in
combination with the Siegert approach. Data in (a) are taken from~\cite{faul}
and in (b) from~\cite{faul} (circles) and~\cite{berman} (pluses).
}
\label{figFT}
\end{figure}

\end{document}